\documentclass{article}
\usepackage{spconf,epsfig,cite}
\usepackage{amsmath,graphicx,subfig}
\usepackage{tikz}
\usetikzlibrary{decorations.pathreplacing}
\usetikzlibrary{chains,arrows,shapes,spy}
\usepackage{caption}
\usepackage{enumitem}
\tikzset{
	font={\fontsize{9}{11.0476pt}\selectfont}}
\usepackage{mathtools,amssymb,cuted}
\usepackage{pgfplots}
\pgfplotsset{compat=newest}
\usepackage{romannum}

\usepackage{algpseudocode}
\usepackage{algorithm}
\algnewcommand{\Parameters}[1]{%
	\State \textbf{\underline{Parameters}:}
	\Statex \hspace*{\algorithmicindent}\parbox[t]{.8\linewidth}{\raggedright #1}
}

\algnewcommand{\Input}[1]{%
	\State \textbf{\underline{Inputs}:}
	\Statex \hspace*{\algorithmicindent}\parbox[t]{.8\linewidth}{\raggedright #1}
}

\algnewcommand{\Initialization}[1]{%
	\State \textbf{\underline{Initialization}:}
	\Statex \hspace*{\algorithmicindent}\parbox[t]{.8\linewidth}{\raggedright #1}
}

\algnewcommand{\Iteration}[1]{%
	\State \textbf{\underline{Iteration}:}
	\Statex \hspace*{\algorithmicindent}\parbox[t]{.8\linewidth}{\raggedright #1}
}

\title{Zero-Crossing Precoding with Maximum Distance to the Decision Threshold for Channels With 1-Bit Quantization and Oversampling}

\name{Diana~M.~V.~Melo,~Lukas~T.~N.~Landau~and~Rodrigo~C.~de~Lamare}
\address{Centre for Telecommunications Studies\\
	Pontifical Catholic University of Rio de Janeiro, 
	Rio de Janeiro, Brazil 22453-900\\
	Email: diana;lukas.landau;delamare@cetuc.puc-rio.br}

\begin{document}
\ninept

\maketitle

\begin{abstract}
Low-resolution devices are promising for systems that demand low energy consumption and low complexity as required in IoT systems. In this study, we propose a novel waveform for bandlimited channels with 1-bit quantization and oversampling at the receivers. The proposed method implies that the information is conveyed within the time instances of zero-crossings which is then utilized in combination with a
 Gray-coding scheme.  
Unlike the existing method, the proposed method does not require optimization and transmission of a dynamic codebook.
The proposed approach outperforms the state-of-the-art method in terms of bit error rate.
\\
\end{abstract}

\vspace{-1em}
\section{Introduction}

Internet of things (IoT) is an integrated part of future communication networks whose most promising applications include smart grid, intelligent transportation and industrial automation \cite{Chen_2019,Gupta_2015}. In terms of the energy consumption and complexity of IoT devises, low resolution analog-to-digital converters (ADCs) are promising. Hence, there is great interest in the development of communication systems with 1-bit receivers. 
Established figures of merit suggest that achieving resolution in time is less challenging than resolution in amplitude, e.g., \cite{Walden_1999}.
With this, it is promising to balance the loss of information caused by the coarse quantization by oversampling in time.
In \cite{shamai_1994}
a significant gain for oversampling in terms of the achievable rate is shown
 by the consideration of Zakai bandlimited processes, which convey the information within the time instances of zero-crossings. 
Later a significant gain from oversampling in the achievable rate is shown for the noisy bandlimited channel \cite{Landau_CL2017}.
A more practical approach was presented in \cite{Landaus_2013}, where filter coefficients are optimized for a 16-QAM transmission,
in terms of the maximization of the minimum distance to the decision threshold, which is widely used also in other studies with quantization \cite{LandauDel_2017,Mo_2015, Jedda_2018}.
Later the approach in \cite{Landaus_2013} was improved and extended for the multiple-input-single-output channel \cite{Gokceoglu_2016} and a more advanced waveform design was proposed for the multiuser multiple-input multiple-output (MIMO) downlink in \cite{LarsonB_2017}, with a separate precoding for time and space.


In this context, our study  
also considers the multiuser MIMO downlink
with 1-bit quantization and oversampling at the receivers with separate precoding in space and time \cite{LarsonB_2017}.
 Unlike the existing method \cite{LarsonB_2017}, the proposed temporal waveform design relies on the information conveyed by the time instances of zero-crossing of the received signal, which is beneficial due to a reduced maximum and average number of zero-crossings per time interval. In the presence of a design restriction by bandlimitation, a desired output pattern with less zero-crossings is promising for the waveform optimization. 
  Different to the existing method \cite{LarsonB_2017} precoding and detection process are no longer symbol-by-symbol operations and rely on the sign of the previous encoded symbol and received symbol, respectively.  
Moreover, a Gray code is devised for the zero-crossing modulation. The proposed waveform is presented in a flexible fashion with variable signaling rate including faster-than-Nyquist signaling and variable sampling rate. 
Numerical results show a superior performance in terms of BER in comparison to the state-of-the-art method \cite{LarsonB_2017}.


The rest of the study is organized as follows: Section~\ref{sec:system_model} describes the system model,  
whereas Section~\ref{sec:ZF} details the spatial precoding.
Section~\ref{sec:PoP} details the proposed zero-crossing (ZC) temporal precoding design, the corresponding optimization problem and the detection process as well as depicts the Gray-code for zero-crossing modulation. Section~\ref{sec:numerical_results} presents and discusses numerical results, while Section~\ref{sec:conclusiones} gives the conclusions.

Notation: In the paper all scalar values, vectors and matrices are represented by: ${a}$, ${\boldsymbol{x}}$ and ${\boldsymbol{X}}$, respectively. We use the tilde sign on the top when referring to complex quantities. 

%

\section{System Model}
\label{sec:system_model}
\begin{figure}[h]
\begin{center}
\captionsetup{justification=centering}
\includegraphics[width=9cm]{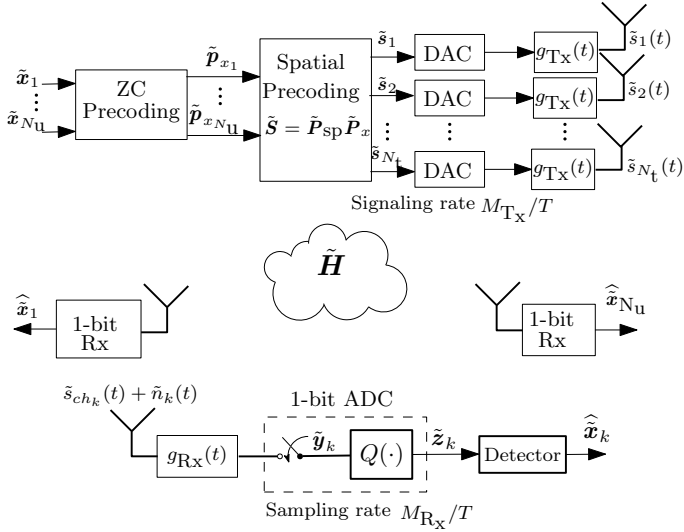}
\caption{Multiuser MIMO system model}
\label{fig:system_mimomodel}       
\vspace{-0.8em}
\end{center}
\vspace{-0.6em}
\end{figure}

We consider a multiuser MIMO system,
as shown in
Fig.~\ref{fig:system_mimomodel},
with $N_{\text{u}}$ single antenna users and $N_{\text{t}}$ transmit antennas at the base station (BS).
For each user $k$ a sequence $\tilde{\boldsymbol{x}}_k$, of $N$ complex symbols, each one denoted as $\tilde{x}_{k,i} = x_{k,i}^{I} + x_{k,i}^{Q}$ and associated to the symbol duration $T$, is temporally precoded into the samples vector $\tilde{\boldsymbol{p}}_{\text{x}_{k}}$with length $N_{\text{q}}$, where $N_{\text{q}} = M_{\text{Tx}}N+1$.
Subsequently the temporally precoded sequences are stacked, linear spatially precoded with matrix $\tilde{\boldsymbol{P}}_{\text{sp}}$
with dimensions $N_{\text{t}} \times N_{\text{u}}$
and converted to a continuous waveform with ideal digital-to-analog converter (DAC) with signaling rate $\frac{M_{\text{Tx}}}{T}$ and the pulse shaping filter with impulse response $g_{\text{Tx}}(t)$. At each user the received signal is sampled with sampling rate $\frac{M_{\text{Rx}}}{T}=\frac{ M M_{\text{Tx}}}{T}$ and quantized after passing the receive filter with impulse response $g_{\text{Rx}}(t)$.
A flat fading channel is considered, which is described by the channel matrix  $\tilde{\boldsymbol{H}}$ with dimensions $N_{\text{u}} \times N_{\text{t}} $.  
By stacking $N_{\text{u}}$ users in the first and $N_{\text{tot}}=M_{\text{Rx}}N+1$ time instances in the second dimension,
the received signal can be expressed by a matrix with dimensions $  N_{\text{u}} \times  N_{\text{tot}}$ described by
\begin{align}
\label{eq:received_signal}
\tilde{\boldsymbol{Z}}= \mathrm{Q}(    \tilde{\boldsymbol{H}}  \tilde{\boldsymbol{P}}_{\textrm{sp}} \tilde{\boldsymbol{P}}_{\textrm{x}}+\tilde{\boldsymbol{N}}  \boldsymbol{G}_{\text{Rx}}^T  ) \text{,}
\end{align}
where $Q(\cdot)$ denotes the element-wise applied quantization operator due to the 1-bit ADC.
The combined waveform determined by the transmit and receive filter $ v (t) = g_{\text{Tx}}(t) * g_{\text{Rx}}(t) $ is taken into account in the matrix  
$\tilde{\boldsymbol{P}}_{\textrm{x}}=
  \left [ (\boldsymbol{V}
 \boldsymbol{U}
 {\tilde{\boldsymbol{p}}}_{\text{x}_{1}})^{\text{T}};(\boldsymbol{V} \boldsymbol{U} \tilde{{\boldsymbol{p}}}_{\text{x}_{2}})^{\text{T}}; \ldots ; (\boldsymbol{V}
 \boldsymbol{U}
 \tilde{{\boldsymbol{p}}}_{\text{x}_{N_{\text{u}}}})^{\text{T}}\right ]$, where the waveform matrix with dimensions $N_{\text{tot}} \times N_{\text{tot}}$  reads as
 \begin{align}
\scriptsize{
\label{eq:MatrixV}
  \mathbb{\boldsymbol{V}} = \;
   \begin{bmatrix}
      v\left ( 0 \right ) & v\left ( \frac{T}{M_{\text{Rx}}} \right ) & \cdots &  v\left (T N \right ) \\
      v\left ( -\frac{T}{M_{\text{Rx}}} \right ) & v\left ( 0 \right ) & \cdots &  v\left (T \left ( N-\frac{1}{M_{\text{Rx}}} \right ) \right ) \\
      \vdots & \vdots & \ddots & \vdots \\
			v\left (-T N \right ) &  v\left (T \left ( -N+\frac{1}{M_{\text{Rx}}} \right ) \right ) & \cdots &  v\left ( 0 \right )
   \end{bmatrix} \text{.}
}
\end{align}
The matrix $\boldsymbol{U}$ with dimensions  $N_{\text{tot}} \times N_{\text{q}}$ describes the $M$-fold upsampling operation and is defined by
{\begin{align}
\label{eq:Matrizu}
\boldsymbol{U}_{m,n}=
\begin{cases}
  1,  & \textrm{for} \quad m = M \cdot \left ( n-1 \right )+1\\
  0, &  {\textrm{else.}}
\end{cases}
\end{align}}
The matrix $\tilde{\boldsymbol{N}}$
with dimensions $N_{\text{u}} \times 3N_{\text{tot}} $ contains i.i.d.\ complex Gaussian noise samples with zero mean and variance $\sigma^{2}_{\tilde{n}}$. 
The matrix $\boldsymbol{G}_{\textrm{Rx}}$ represents the receive filter and reads as
\begin{align}
\label{eq:GRx}
\boldsymbol{G}_{\textrm{Rx}}=  a_{\textrm{Rx}} \begin{bmatrix}
\left[ \ \boldsymbol{g}_{\text{Rx}}^T \ \right]  \ 0 \cdots \ \ \ 0  \\
0 \ \left[ \ \boldsymbol{g}_{\text{Rx}}^T \ \right] \ 0 \cdots 0 \\
\ddots  \ddots \ddots    \\
0 \cdots \ \ \  0 \ \left[ \ \boldsymbol{g}_{\text{Rx}}^T  \ \right]   
\end{bmatrix}_{N_{\text{tot}}  \times 3N_{\text{tot}} }
\text{,}
\end{align} 
with
$\boldsymbol{g}_{\text{Rx}} = [ g_{\text{Rx}} (-T  ( N + \frac{1}{M_{\text{Rx}}}  )  ),
g_{\text{Rx}}(-T  ( N + \frac{1}{M_{\text{Rx}}} )  + \frac{T}{M_{\text{Rx}}}  ), \ldots,$
$ g_{\text{Rx}} (T ( N + \frac{1}{M_{\text{Rx}}} )  ) ]^T $ and $a_{\textrm{Rx}}=  (T / ( M_{\text{Rx}} ))^{1/2} $.
The next section describes the spatial precoder as suggested in \cite{LarsonB_2017}.

\section{Spatial Zero-forcing Precoding}
\label{sec:ZF}
Assuming perfect channel state information the spatial zero-forcing (ZF) precoding matrix \cite{SpencerHaardt_2004} is described by
\begin{equation}
\label{eq:zf}
     \tilde{\boldsymbol{P}}_{\text{sp}}=
     \tilde{\boldsymbol{P}}_{\text{zf}} = c_{\text{zf}}\tilde{\boldsymbol{P}'}_{\text{zf}}
     \quad \textrm{with} \quad
     \tilde{\boldsymbol{P}'}_{\text{zf}} = \tilde{\boldsymbol{H}}^{\text{H}}\left ( \tilde{\boldsymbol{H}}\tilde{\boldsymbol{H}}^{\text{H}} \right )^{-1}   \textrm{.}
\end{equation} 
 The temporal transmit signals are described in the matrix 
 $\tilde{{\boldsymbol{P}}}_{\text{x,Tx}} = \left [ (\boldsymbol{G}^{{T}}_{\text{Tx}}
 \boldsymbol{U}
 {\tilde{\boldsymbol{p}}}_{\text{x}_{1}})^{{T}};(\boldsymbol{G}^{{T}}_{\text{Tx}}
 \boldsymbol{U}
 {\tilde{\boldsymbol{p}}}_{\text{x}_{2}})^{{T}}; \ldots ; (\boldsymbol{G}^{{T}}_{\text{Tx}}
 \boldsymbol{U}
 {\tilde{\boldsymbol{p}}}_{\text{x}_{N_{\text{u}}}})^{{T}}\right ]$, with the Toeplitz matrix
 \begin{align}
\label{eq:GTx}
\boldsymbol{G}_{\textrm{Tx}}=  a_{\textrm{Tx}} \begin{bmatrix}
\left[ \ \boldsymbol{g}_{\text{Tx}}^T \ \right]  \ 0 \cdots \ \ \ 0  \\
0 \ \left[ \ \boldsymbol{g}_{\text{Tx}}^T \ \right] \ 0 \cdots 0 \\
\ddots  \ddots \ddots    \\
0 \cdots \ \ \  0 \ \left[ \ \boldsymbol{g}_{\text{Tx}}^T  \ \right]   
\end{bmatrix}_{N_{\text{tot}}  \times 3N_{\text{tot}} }
\text{,}
\end{align} 
with
$\boldsymbol{g}_{\text{Tx}} = [ g_{\text{Tx}} (-T  ( N + \frac{1}{M_{\text{Rx}}}  )  ),
g_{\text{Tx}}(-T  ( N + \frac{1}{M_{\text{Rx}}} )  + \frac{T}{M_{\text{Rx}}}  ), \ldots,$
$ g_{\text{Tx}} (T ( N + \frac{1}{M_{\text{Rx}}} )  ) ]^T $ and $a_{\textrm{Tx}}=  (T / ( M_{\text{Rx}} ))^{1/2} $.
 With this and by considering uncorrelated user signals and equal power allocation for all users in terms of $E_{\text{Tx}}$, a signal covariance matrix can be defined as ${\boldsymbol{R}} = \mathbb{E}\left \{ {\tilde{\boldsymbol{P}}}_{\text{x,Tx}}{\tilde{\boldsymbol{P}}}_{\text{x,Tx}}^{\text{H}} \right \}  = E_{\text{Tx}} \boldsymbol{I}_{N_\text{u}}$. 
Including the spatial precoding, the total transmit energy can be cast as $E_{\text{0}}= \text{trace}\left (\tilde{\boldsymbol{P}}_{\text{zf}}{\boldsymbol{R}} \tilde{\boldsymbol{P}}_{\text{zf}}^{\text{H}} \right ) $.
Using $\frac{E_{\text{0}}}{E_{\text{Tx}}}= N_{\text{u}}$ yields the ZF scaling factor 
\begin{align}
c_{\text{zf}} =  \left (  N_{\text{u}}/ \mathrm{trace} \left (  \left ( \tilde{\boldsymbol{H}}\tilde{\boldsymbol{H}}^{\text{H}} \right )^{-1} \right )  \right )^{\frac{1}{2}} \text{.}
\end{align}
For ZF precoding, $\tilde{\boldsymbol{H}}\tilde{\boldsymbol{P}}_{\text{sp}}$ becomes a diagonal matrix $\beta \, \boldsymbol{I}_{N_{\text{u}}} $ where $\beta$ refers to the real valued beamforming gain, which is equal to $c_{\text{zf}}$. 
With this, \eqref{eq:received_signal}
simplifies to
\begin{align}
\tilde{\boldsymbol{Z}}= \mathrm{Q}(    \beta \tilde{\boldsymbol{P}}_{\textrm{x}}+\tilde{\boldsymbol{N}}  \boldsymbol{G}_{\text{Rx}}^T  ) \text{,}
\end{align}
and the signal at the $k$-th user can be written as a column vector as
\begin{align}
\tilde{\boldsymbol{z}}_{k}=Q(
\beta \boldsymbol{V} \boldsymbol{U} \tilde{\boldsymbol{p}}_{\textrm{x}_{k}}+\boldsymbol{G}_{\textrm{Rx}} \tilde{\boldsymbol{n}}_k  )\text{,}
\end{align}
where the vector $\tilde{\boldsymbol{n}}_k$ 
with length
$3N_{\text{tot}}$ contains i.i.d.\ complex Gaussian noise samples with zero mean and variance $\sigma^{2}_{\tilde{n}}$.

In the sequel it is considered that the real and imaginary part
can be processed independently for temporal precoding and detection and then only a real valued process is studied, denoted by $\boldsymbol{p}_{\text{x}_{k}}$ without the tilde and with block transmit energy $E_{\text{Tx}} / 2= E_{\text{0}}/ (2N_{\text{u}})$.
For improving the readability we drop the index $k$ in the sequel. The next section describes the proposed method for the construction of $\boldsymbol{p}_{\text{x}}$ based on the input symbol sequence $\boldsymbol{x}$.

\section{Proposed Zero-Crossing Precoding}
\label{sec:PoP}
\begin{figure*}[t] 
\begin{minipage}{5cm} 
\begin{center}
\includegraphics[width=13cm]{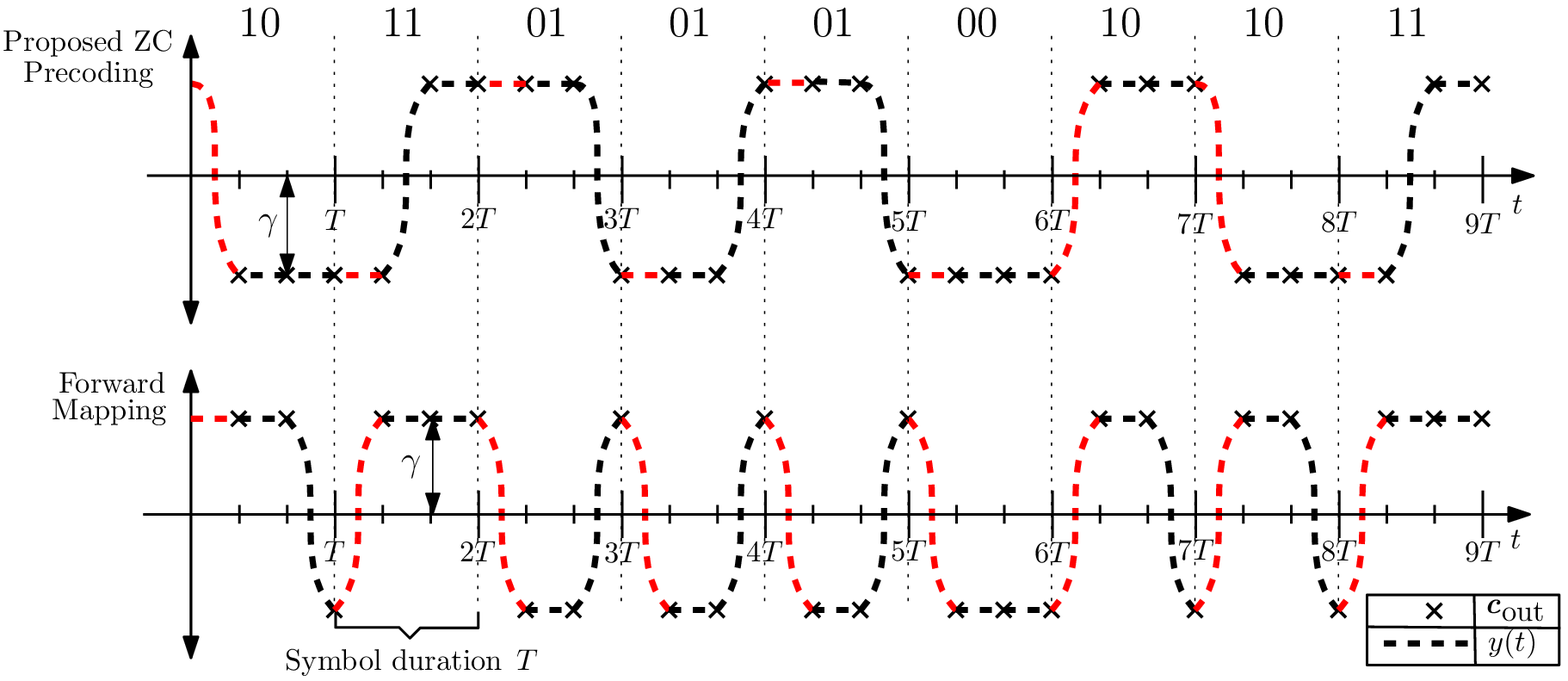}
\end{center}
\end{minipage} 
\hspace{0.45\linewidth}
\begin{minipage}{5cm} 
\vspace{-0.3\linewidth}
\includegraphics[width=5.3cm]{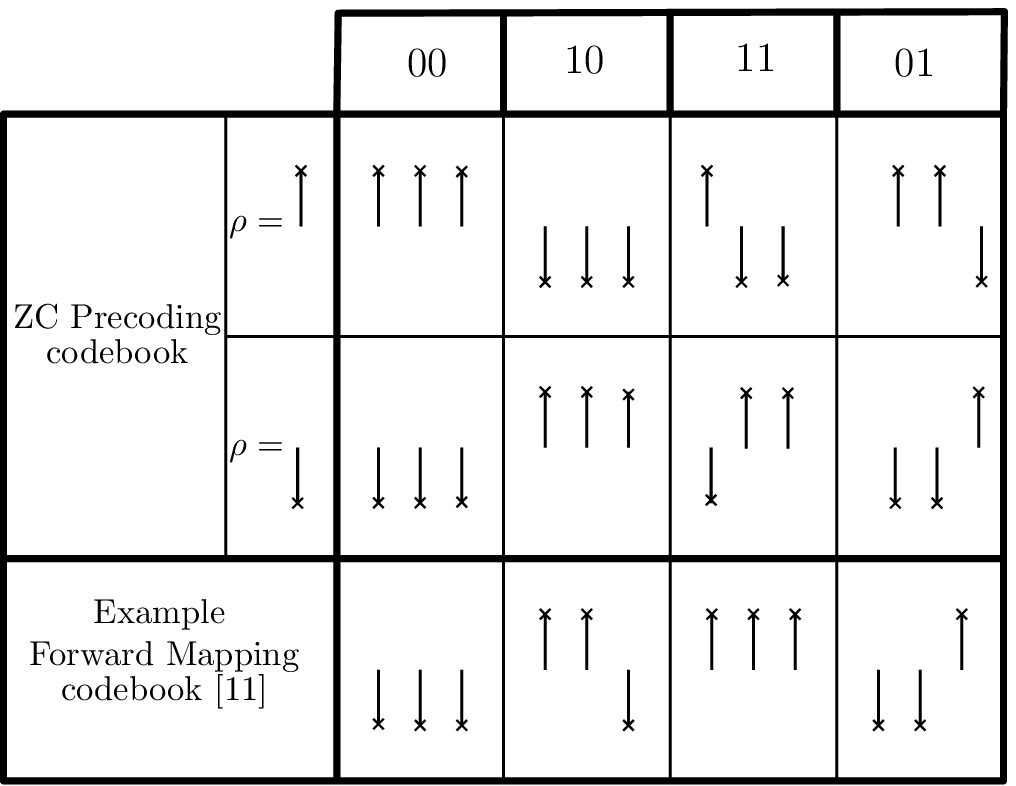}
\end{minipage} 
\caption{Mapping process for   construction of $\boldsymbol{c}_{\text{out}}$ for ZC precoding and an example mapping for Q-Precoding \cite{LarsonB_2017}, $M_{\textrm{Rx}}=3$}
\label{fig:illustration}
\end{figure*} 
In this section a precoding method is proposed, which conveys the information in the zero-crossing time instances of the received signal.
In the presented channel with oversampling at the receiver, a symbol interval has $M_{\textrm{Rx}}$ samples. In other words, the symbol interval is subdivided into $M_{\textrm{Rx}}$ sub intervals. 
The principal idea of the proposed ZC precoding is that a transmit symbol
appears in terms of a zero-crossing in one of the $M_{\textrm{Rx}}$ sub intervals or in terms of the absence of a zero-crossing.
With this, the input cardinality is
$R_{\text{in}}=M_{\textrm{Rx}}+1$ per real valued dimension.
Then each symbol conveys
$\log_2 (M_{\textrm{Rx}} + 1 )$ bits, per real dimension.
According to the study in \cite{shamai_1994} with Zakai bandlimited processes, it is possible to construct a bandlimited process even with zero-crossings in each Nyquist interval. In this context the present approach is promising as only 
$M_{\textrm{Rx}}$ out of $M_{\textrm{Rx}} + 1$ symbols imply a zero-crossing. 
The desired binary output pattern for a sequence of $N$ symbols, which yields the zero-crossings in the desired intervals, shall be denoted as $\boldsymbol{c}_{\text{out}}$.  
The output pattern can be constructed sequentially by concatenation of sequence segments $\boldsymbol{c}_{\textrm{s},i}$ of $M_{\textrm{Rx}}$ samples, which depends on the input symbol ${x}_{i} \in  \mathcal{X}_{\text{in}}:=  \left \{ b_{1},b_{2} \cdots , b_{R_{\text{in}}}\right \}$, from the sequence ${\boldsymbol{x}}=[x_0,\ldots,x_i,\ldots, x_{N-1}]$ and also on the last sample of the previous sequence segment termed $\rho_{i-1}$. 
An example for the construction of $\boldsymbol{c}_{\textrm{out}}$ 
with $M_{\textrm{Tx}}=3$
is shown in Fig.~\ref{fig:illustration} together with the forward mapping (FM) approach known from literature \cite{LarsonB_2017}.  
As mentioned, the zero-crossing mapping implies the condition on the last sample $\rho_{i-1} \in \{1,-1\}$ of the previous code segment $\boldsymbol{c}_{\textrm{s},i-1}$ such that each input symbol is associated with two possible code segments. The pattern sequence $\boldsymbol{c}_{\text{out}}=[pb,\boldsymbol{c}_{\textrm{s,0}}^T,\ldots,
\boldsymbol{c}_{\textrm{s,N-1}}^T
]^T$
with the length $N M_{\textrm{Rx}}+1$
starts with a pilot sample $pb$, which initiates the sequence construction and later the detection process.
Note that FM theoretically supports a larger input alphabet cardinality of $2^{M_{\textrm{Rx}}}$, which however involves an increased number of zero-crossings, more difficult to construct in the presence of bandlimitation,
cf.\
\cite{LarsonB_2017}.

The next section explains the construction of a transmit sequence ${\boldsymbol{p}}_{\text{x}}$, for inducing the desired output pattern $\boldsymbol{c}_{\text{out}}$.
\subsection{Convex Optimization for Zero-Crossing Precoding}
Generally, samples close to the decision threshold are more sensitive to noise and with this, it is promising to develop a precoder that maximizes the minimum distance, termed $\gamma$. 
Stemming from above, in this section the optimization process is described to obtain the optimal temporal precoding vector ${\boldsymbol{p}}_{\text{x}}$.
The corresponding equivalent optimization problem, similar to the ones in \cite{LarsonB_2017,Landau_SCC2013}, can be expressed in the epigraph form, cf.\ Section 4.1.3 in \cite{Boyd_2004},  with 
\begin{equation}
\label{eq:convex}
\begin{aligned}
& \min_{\boldsymbol{r}}
& & \boldsymbol{a}^{\text{T}}\boldsymbol{r} \\
& \text{subject to:}
& & \boldsymbol{B}\boldsymbol{r} \leq 0 \\
&&&  \boldsymbol{r}^{\text{T}}   \boldsymbol{W} ^{\text{T}} \boldsymbol{W}\boldsymbol{r}  \leq 
 \frac{1}{2} {E_{\text{Tx}}}
\text{,}
\end{aligned}
\end{equation}
where the optimization variable contains the temporal precoding vector and the distance to the decision threshold with negative sign in terms of $\boldsymbol{r}   = \left [ \boldsymbol{p}_{\text{x}}^T,-\gamma \right ]^{\text{T}}$. 
Together with 
$\boldsymbol{a} = \left [ \boldsymbol{0}_{1\times N_{\text{q}}},1 \right]^{\text{T}}$
and the first constraint with
 $\boldsymbol{B} = -\left [ \beta \boldsymbol{C}\boldsymbol{V}\boldsymbol{U}, \boldsymbol{1}_{N_{\text{tot}}\times 1} \right]$ and
$\boldsymbol{C} = \textrm{diag}\left (  \boldsymbol{c}_{\text{out}}  \right )$ the optimization problem maximizes the minimum distance to the decision threshold in the desired direction induced by $\boldsymbol{c}_{\text{out}}$.
The second constraint with
$ \boldsymbol{W} = \left [  \boldsymbol{G}^{T}_{{\text{Tx}}}\boldsymbol{U} , \boldsymbol{0}_{N_{\text{tot}}\times 1} \right]$ accounts for the transmit energy constraint.
Note that, different to the problem formulation in \cite{LarsonB_2017}, the energy constraint in \eqref{eq:convex} explicitly takes into account $g_{\textrm{Tx}}(t)$ as the sequence $\boldsymbol{p}_{\text{x}}$ does not have a white power spectrum for the considered cases.
The described problem \eqref{eq:convex} is a convex quadratically constrained
quadratic program, since $\boldsymbol{W}^{\text{T}}\boldsymbol{W} \in S_{+}^{n}$  (cf. Section 4.4 in \cite{Boyd_2004}).

Implicitly, the optimization problem shapes the waveform $y(t)$ at the receiver, which is described in the discrete model by $\beta \boldsymbol{V} \boldsymbol{U} \boldsymbol{p}_{\textrm{x}}$ for the noiseless case. 
To highlight the difference between the mapping strategies
Fig.~\ref{fig:illustration} shows a noncritical example, where transmit and receive filter still provide sufficient bandwidth, such that $\gamma$ is only limited by the energy constraint. In Fig.~\ref{fig:illustration} it can be seen that the proposed zero-crossing sequence $\boldsymbol{c}_{\text{out}}$ involves significantly less zero-crossings than a sequence constructed by an equivalent forward mapping \cite{LarsonB_2017}. Numerical results in Section~\ref{sec:numerical_results}
will show the impact of bandlimitation on the optimization problem in terms of $\gamma$ for the both mapping strategies.
\subsection{Detection}
\begin{table}[b]
\caption{Gray Code for $M_{\text{Rx}}=2$}
\label{tab:GC2} \vspace{-2em}
\begin{center}
\scalebox{0.67}{
\begin{tabular}{|l|c|c|c|c|c|c|c|c|}
\hline
Gray Code         & $000$ & $001$ & $011$ & $010$ & $110$ & $111$ & $101$ & $100$  \\ \hline
$ \rho_{2i-1}=1$  &       &       &       &       &       &       &     &     \\ 
$ [\boldsymbol{c}_{\textrm{s},2i}^T, \boldsymbol{c}_{\textrm{s},2i+1}^T ]$
& 1111 & 111-1 & 11-1-1 & 1-1-1-1 & 1-1-11 & -1-1-11 & -1-1-1-1 & -1-111 \\ \hline
$ \rho_{2i-1}=-1$  &       &       &       &       &       &       &     &     \\ 
$ [\boldsymbol{c}_{\textrm{s},2i}^T, \boldsymbol{c}_{\textrm{s},2i+1}^T ]$
& -1-1-1-1 & -1-1-11 & -1-111 & -1111 & -111-1 & 111-1 & 1111 & 11-1-1 \\ \hline
\end{tabular}}
\vspace{-0.9em}
\end{center}
\end{table}

Optimal detection like maximum likelihood (ML) detection appears impractical because of the large number of candidates and also because each evaluation of the ML metric involves solving \eqref{eq:convex}.
Moreover, in order to meet the requirements of low complexity devices a simple detection process is favorable. 

Given that the detector shall detect the time instance of the zero-crossing in each symbol interval, its decision also depends on $\rho_{i-1}$. 
With this, based on the received signal \eqref{eq:received_signal}, the detector considers the $\boldsymbol{z}_{b,i}$ sub-sequences each with length  $ M_{\text{Rx}}+1$, where the first sample of  $\boldsymbol{z}_{b,i}$ corresponds to $\hat{\rho}_{i-1}$, required to perform the backward mapping process $\overleftarrow{d}:\boldsymbol{z}_{b,i}\rightarrow \mathcal{X}_{\text{in}}$, where $\boldsymbol{z}_{b,i} \subseteq  \mathcal{C}_{\text{map}}$ and $\mathcal{C}_{\text{map}}$ contains the valid codewords in the form $ \left [ \rho, \boldsymbol{c}_{s}\right ] $. The first symbol in the sequence is detected by taking into account the pilot sample $pb$.
While in the noiseless environment, it is possible to perform a perfect recovery of $\boldsymbol{x}$ only by considering the backward mapping, the presence of noise can alter received samples in such a way such that invalid codewords occur.
 For this cases, additional decision rules are defined in terms of consideration of the Hamming-distance metric \cite{LarsonB_2017}. This means, if $\boldsymbol{z}_{b,i} \nsubseteq  \boldsymbol{c}_{\text{map}}$ the $i$-th symbol is detected according to
\begin{equation*}
\begin{aligned} \widehat{x_{i}} = \overleftarrow{d} \left ( \boldsymbol{c} \right )\text{,} \quad \textrm{where} \quad \boldsymbol{c} =
& arg \; \underset{\boldsymbol{c}_{\text{map}}}{\text{min}}
& & \text{Hamming} \left (\boldsymbol{z}_{b,i}, \boldsymbol{c}_{\text{map}} \right ) \text{,}\\
\end{aligned}
\end{equation*}
where $\text{Hamming} \; ( \boldsymbol{z}_{b,i},\boldsymbol{c}_{\text{map}}) = \sum_{n=1}^{M_{\text{Rx}}+1}\frac{1}{2}\left | \boldsymbol{z}_{b,i,n}- \boldsymbol{c}_{\text{map},n} \right |$.

\subsection{Gray Coding for Zero-Crossing Precoding}
Given a sequence of data bits, a Gray code is applied for the zero-crossing symbols, as shown on the RHS of Fig.~\ref{fig:illustration}. In this context, symbols that have near or consecutive zero-crossings differ only in one data bit.
For cases where the input cardinality $ R_{\text{in}} = M_{\text{Rx}} + 1 $ is not a power of 2, symbol sequences are considered for mapping.
For $R_{\text{in}}=3$, 3 bits are mapped to a sequence of 2 symbols, as shown in Table~\ref{tab:GC2}, which implies a conversion loss of $(1.5-\log_2 3 ) \approx 0.085 \ \text{bits per symbol}$.

\section{Numerical Results}
\label{sec:numerical_results}
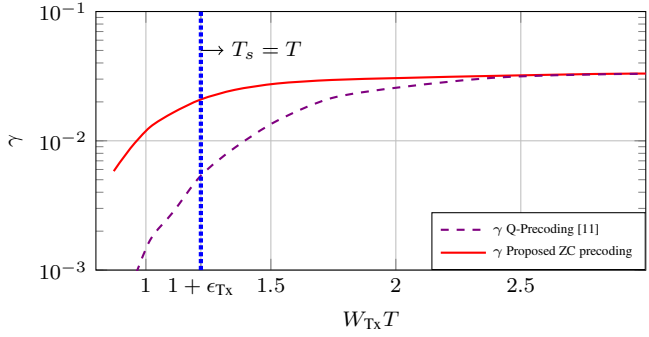
\begin{figure}[t]
\begin{center}
%
%
%


\definecolor{mycolor1}{rgb}{0.00000,1.00000,1.00000}%
\definecolor{mycolor2}{rgb}{1.00000,0.00000,1.00000}%

\pgfplotsset{every axis label/.append style={font=\footnotesize},
every tick label/.append style={font=\footnotesize},
every plot/.append style={ultra thick} 
}

\begin{tikzpicture}[font=\footnotesize] 

\begin{axis}[%
name=ber,
ymode=log,
width  = 0.85\columnwidth,
height = 0.4\columnwidth,
scale only axis,
xmin  = 0.8,
xmax  = 3,
xlabel= {$W_{\text{Tx}}T$},
xmajorgrids,
ymin=0.001,
ymax=0.1,
ylabel={$\gamma$},
ymajorgrids,
legend entries={$\gamma$ Q-Precoding \cite{LarsonB_2017},
                $\gamma$ Proposed ZC precoding,	
														},
xtick={1,1.22 ,1.5,2,2.5},
        xticklabels={$1$,$1+\epsilon_{\text{Tx}}$  , $1.5$,$2$,$2.5$},
legend style={at={(1,0)},anchor=south east,draw=black,fill=white,legend cell align=left,font=\tiny}
]

\addlegendimage{smooth,color=violet,dashed, thick, mark=none,
y filter/.code={\pgfmathparse{\pgfmathresult-0}\pgfmathresult}}
\addlegendimage{smooth,color=red,solid, thick, mark=none,
y filter/.code={\pgfmathparse{\pgfmathresult-0}\pgfmathresult}}

\addplot+[smooth,color=red,solid, thick, every mark/.append style={solid, fill=gray!20} ,mark=none,
y filter/.code={\pgfmathparse{\pgfmathresult-0}\pgfmathresult}]
  table[row sep=crcr]{%
0.871428571428571	        0.00582339998635118\\
0.938461538461539	        0.00877288024332817\\
1.01666666666667	        0.0127664936505643\\
1.10909090909091	        0.0165263736175289\\
1.22	                    0.0208952967947965\\
1.35555555555556	        0.0247848190362989\\
1.525                   	0.0277399032459994\\
1.74285714285714        	0.0295491675042235\\
2.03333333333333        	0.0306458863277253\\
2.44                    	0.0319171659810108\\
3.05                    	0.0331068757127887\\
4.06666666666667        	0.0310387680242437\\
6.1                     	0.0315616771216149\\
12.2                    	0.0315907093065039\\
};

\addplot+[smooth,color=violet,dashed, thick, every mark/.append style={solid, fill=gray!20},mark=none,
y filter/.code={\pgfmathparse{\pgfmathresult-0}\pgfmathresult}]
  table[row sep=crcr]{%
0.871428571428571   	0.000449478239979334\\
0.938461538461539	    0.000744938926845561\\
1.01666666666667    	0.00170360475066285\\
1.10909090909091    	0.00281256791154467\\
1.22                	0.0053764155906486\\
1.35555555555556	    0.00883922207723862\\
1.525	                0.0142932284578272\\
1.74285714285714    	0.0213989099339389\\
2.03333333333333	    0.0261740631887097\\
2.44	                0.0311252406148608\\
3.05                	0.0330471560798835\\
4.06666666666667	    0.0323566158531291\\
6.1	                    0.0317904750933107\\
12.2                	0.0315108268967575\\
};

\draw[->] (1.22, 0.05) -- (1.31, 0.05);
\draw [densely dotted, ultra thick, blue] (1.22,0.001) -- (1.22,0.1);
\node[anchor=west] at (1.31, 0.05) (text) {$T_{s}=T$};

\end{axis}

%

\end{tikzpicture}%
\caption{$\gamma$ vs bandwidth   for $M_{\text{Rx}}=M_{\text{Tx}} = 2$} 
\label{fig:gamma}       
\vspace{-1em}
\end{center}
\end{figure}
In all the experiments a raised-cosine (RC) pulse shaping filter with roll-off factor $\epsilon_{\text{Tx}} = 0.22$ is considered, while the receive filter is a square-root raised cosine (RRC) with roll-off factor $\epsilon_{\text{Rx}} = \epsilon_{\text{Tx}} = 0.22$.
The Tx and Rx filter bandwidth is $W_{\text{Rx}}=W_{\text{Tx}} =\left ( 1 + \epsilon_{\text{Tx}} \right )/T_{\text{s}}$. The influence of the bandwidth symbol duration product on  $\gamma$ is
shown in Fig.~\eqref{fig:gamma}, which confirms that with large bandwidth, gamma is only restricted by the transmit energy and with decreasing bandwidth, $\gamma$ tends to zero. As expected, the ZC precoding provides a larger distance to the decision threshold when reducing bandwidth as compared to the forward mapping in \cite{LarsonB_2017}.
The case for $W_{\text{Tx}}T=1$ supports the theory that $\log_2(M_{\text{Rx}}+1)$ bits per Nyquist interval can be transmitted in the noise-free case \cite{shamai_1994}, and that there is a reasonable tolerance against noise. 
For the residual simulations the common setting $T_{\textrm{s}}=T$ is considered.
The numerical evaluation the uncoded BER is considered, where the signal-to-noise ratio is defined as
\begin{align}
\text{SNR} = \frac{   E_{\text{0}} / (N_{\text{q}}T)  } {{N_{0}}\left ( 1+\epsilon_{\text{Rx}} \right )\frac{1}{T}} = \frac{E_{\text{0}}}{N_{\text{q}}N_{0} \left ( 1+\epsilon_{\text{Rx}} \right )} \textrm{,}
\end{align}
where $N_{0}$ is the complex noise power density. 
The considered channel can be expressed as $\tilde{\boldsymbol{H}} = {\tilde{\boldsymbol{G}}_{\text{H}}} \boldsymbol{D_{\text{H}}}^{\frac{1}{2}}$, where $\tilde{\boldsymbol{G}}_{\text{H}}$ is the $N_{\text{u}} \times N_{\text{t}}$ matrix of fast fading coefficients described by a Rayleigh probability density function and $\boldsymbol{D_{\text{H}}}$ is a diagonal matrix which models the geometric attenuation and shadow fading defined as $\boldsymbol{D_{\text{H}}} = \text{diag} \left ({\boldsymbol{\zeta}}/{\left ( {d}/{r_{d}} \right )^{\nu}}\right )$
where $\zeta$ is a log-normal random variable with standard deviation $\sigma_{\text{shadow}}$, $d$ corresponds to the distance between the transmitter and the receiver, $r_{d}$  is the cell radius and $\nu$ is the path loss exponent \cite{Larson_2013}.
The channel parameters are: $r_{d} = 1000$ m, $d = 300$ m, $\sigma_{\text{shadow}} = 8$ dB and $\nu = 3$.
The performance of the proposed method is compared with the precoder in \cite{LarsonB_2017} using equal data rates and a bandwidth constraint incorporated only by the pulse shaping filter.
In order to provide the same conditions,
the proposed ZC precoding method is compared with the method in \cite{LarsonB_2017} based on the optimization problem \eqref{eq:convex}, where the sequence $\boldsymbol{c}_{\text{out}}$ is replaced according to the optimized forward mapping scheme devised in \cite{LarsonB_2017}.
Fig.~\ref{fig:BER} shows the BER performance for different $M_{\text{Rx}}$ and $M_{\text{Tx}}$. 
In addition, a common QPSK modulated signal is presented as a reference, corresponding to only 2 bits per time interval $T$. 
In all the considered cases, 
the proposed ZC precoding has a significantly lower BER than the state-of-the-art method \cite{LarsonB_2017}.
Among the considered cases the configuration with $M_{\text{Rx}} = M_{\text{Tx}} = 2$ shows the best performance in BER.
In addition, the power spectral density for the considered waveforms is shown in Fig.~\ref{fig:PSD}.

\begin{figure}[t]
\begin{center}
%
%
%

\definecolor{mycolor1}{rgb}{0.00000,1.00000,1.00000}%
\definecolor{mycolor2}{rgb}{1.00000,0.00000,1.00000}%

\pgfplotsset{every axis label/.append style={font=\footnotesize},
every tick label/.append style={font=\footnotesize},
every plot/.append style={ultra thick} 
}

\begin{tikzpicture}[font=\footnotesize] 

\begin{axis}[%
name=ber,
ymode=log,
width  = 0.85\columnwidth,
height = 0.5\columnwidth,
scale only axis,
xmin  = 0,
xmax  = 10,
xlabel= {SNR  [dB]},
xmajorgrids,
ymin=0.0000152,
ymax=1,
ylabel={BER},
ymajorgrids,
legend entries={Q-Precoding \cite{LarsonB_2017},
                Proposed ZC precoding,
								Conventional QPSK,	
														},
legend style={at={(0,0)},anchor=south west,draw=black,fill=white,legend cell align=left,font=\tiny}
]

\addlegendimage{smooth,color=black,dashed, thick, mark=none,
y filter/.code={\pgfmathparse{\pgfmathresult-0}\pgfmathresult}}
\addlegendimage{smooth,color=black,solid, thick, mark=none,
y filter/.code={\pgfmathparse{\pgfmathresult-0}\pgfmathresult}}
\addlegendimage{smooth,color=black, thick, densely dotted,mark=none,
y filter/.code={\pgfmathparse{\pgfmathresult-0}\pgfmathresult}}

\addplot+[smooth,color=black,densely dotted,thick, every mark/.append style={solid, fill=gray!20} ,mark=none,
y filter/.code={\pgfmathparse{\pgfmathresult-0}\pgfmathresult}]
  table[row sep=crcr]{%
0	   0.0252282352941176\\
2    0.00818901960784314\\
4	   0.00180235294117647\\
6	   0.000222745098039216\\
8	   1.45098039215686e-05\\
};

\addplot+[smooth,color=teal,solid, thick, every mark/.append style={solid, fill=gray!20} ,mark=asterisk,
y filter/.code={\pgfmathparse{\pgfmathresult-0}\pgfmathresult}]
  table[row sep=crcr]{%
0		0.310339483394834\\
2		0.271712177121771\\
4		0.228014760147601\\
6		0.181557195571956\\
8		0.137785977859779\\
10	0.0991365313653137\\
12	0.0623247232472325\\
14	0.0350036900369004\\
16	0.0156457564575646\\
18	0.00606642066420664\\
20	0.0016309963099631\\
};

\addplot+[smooth,color=green,solid, thick, every mark/.append style={solid, fill=gray!20},mark=square,
y filter/.code={\pgfmathparse{\pgfmathresult-0}\pgfmathresult}]
  table[row sep=crcr]{%
0		 	0.122866666666667\\
2	 		0.067696000000000\\
4	 		0.029680000000000\\
6	 		0.009374666666667\\
8	 		0.002074666666667\\
10		2.773333333333333e-04\\
12	  2.266666666666667e-05\\
14	  0\\
16	  0\\
18	  0\\
20	  0\\
};

\addplot+[smooth,color=magenta,solid, thick, every mark/.append style={solid, fill=gray!20},mark=triangle,
y filter/.code={\pgfmathparse{\pgfmathresult-0}\pgfmathresult}]
  table[row sep=crcr]{%
0		 	0.256591466666667\\
2	 		0.193683733333333\\
4	 		0.131802133333333\\
6	 		0.077002133333333\\
8	 		0.036640533333333\\
10		0.013677866666667\\
12	  0.003735466666667\\
14	  7.002666666666667e-04\\
16	  8.213333333333334e-05\\
18	  5.866666666666667e-06\\
};

\addplot+[smooth,color=blue,solid, thick, every mark/.append style={solid, fill=gray!20},mark=+,
y filter/.code={\pgfmathparse{\pgfmathresult-0}\pgfmathresult}]
  table[row sep=crcr]{%
0		 	0.171864000000000\\
2	 		0.123804000000000\\
4	 		0.081092000000000\\
6	 		0.045576000000000\\
8	 		0.021722000000000\\
10		0.007678000000000\\
12	  0.002054000000000\\
14	  3.080000000000000e-04\\
16	  4.800000000000000e-05\\
18	  4.000000000000000e-06\\
};


\addplot+[smooth,color=blue,dashed, thick, every mark/.append style={solid, fill=gray!20},mark=+,
y filter/.code={\pgfmathparse{\pgfmathresult-0}\pgfmathresult}]
  table[row sep=crcr]{%
0			0.290800000000000\\
2			0.265080000000000\\
4			0.245400000000000\\
6			0.214280000000000\\
8			0.192160000000000\\
10		0.162480000000000\\
12		0.131800000000000\\
14		0.105800000000000\\
16		0.083200000000000\\
18		0.063040000000000\\
20		0.044640000000000\\
};

\addplot+[smooth,color=teal,dashed, thick, every mark/.append style={solid, fill=gray!20},mark=asterisk,
y filter/.code={\pgfmathparse{\pgfmathresult-0}\pgfmathresult}]
  table[row sep=crcr]{%
0		0.377439114391144\\
2		0.36819926199262\\
4		0.356678966789668\\
6		0.343859778597786\\
8		0.330147601476015\\
10	0.31819926199262\\
12	0.304206642066421\\
14	0.290110701107011\\
16	0.278332103321033\\
18	0.266214022140221\\
20	0.251557195571956\\
};

\addplot+[smooth,color=green,dashed, thick, every mark/.append style={solid, fill=gray!20},mark=square,
y filter/.code={\pgfmathparse{\pgfmathresult-0}\pgfmathresult}]
  table[row sep=crcr]{%
0		 	0.325425333333333\\
2	 		0.294512000000000\\
4	 		0.263336000000000\\
6	 		0.228934666666667\\
8	 		0.192214666666667\\
10		0.157845333333333\\
12	  0.125576000000000\\
14	  0.095478666666667\\
16	  0.071156000000000\\
18	  0.050610666666667\\
20	  0.034461333333333\\
};

\addplot+[smooth,color=magenta,dashed, thick, every mark/.append style={solid, fill=gray!20},mark=triangle,
y filter/.code={\pgfmathparse{\pgfmathresult-0}\pgfmathresult}]
  table[row sep=crcr]{%
0		 	0.392226133333333\\
2	 		0.378604800000000\\
4	 		0.365537600000000\\
6	 		0.350005333333333\\
8	 		0.334379200000000\\
10		0.317760533333333\\
12	  0.300072533333333\\
14	  0.281862933333333\\
16	  0.262172800000000\\
18	  0.242150400000000\\
20	  0.223651200000000\\
};

\addplot[only marks,color=magenta,solid,mark=triangle,
y filter/.code={\pgfmathparse{\pgfmathresult-0}\pgfmathresult}]
  table[row sep=crcr]{%
	1 2\\
};\label{P30}

\addplot[only marks,color=green,solid,mark=square,
y filter/.code={\pgfmathparse{\pgfmathresult-0}\pgfmathresult}]
  table[row sep=crcr]{%
	1 2\\
};\label{P31}

\addplot[only marks,color=blue,solid,mark=+,
y filter/.code={\pgfmathparse{\pgfmathresult-0}\pgfmathresult}]
  table[row sep=crcr]{%
	1 2\\
};\label{P32}

\addplot[only marks,color=teal,solid,mark=asterisk,
y filter/.code={\pgfmathparse{\pgfmathresult-0}\pgfmathresult}]
  table[row sep=crcr]{%
	1 2\\
};\label{P8}

\node [draw,fill=white,font=\tiny,anchor= south  west] at (axis cs: 0,0.00023) {
\setlength{\tabcolsep}{0.5mm}
\renewcommand{\arraystretch}{.8}
\begin{tabular}{l}
\ref{P30}{$\quad M_{\text{Rx}} = 2$, $M_{\text{Tx}} = 1$}\\
\ref{P31}{$\quad M_{\text{Rx}} = 2$, $M_{\text{Tx}} = 2$}\\
\ref{P8}{$\quad M_{\text{Rx}} = 3$, $M_{\text{Tx}} = 1$}\\
\ref{P32}{$\quad M_{\text{Rx}} = 3$, $M_{\text{Tx}} = 3$}\\
\end{tabular}
};

\end{axis}

\end{tikzpicture}%
\caption{BER versus SNR for different $M_{\text{Rx}}$ and $M_{\text{Tx}}$. Fixed parameters are $N=50$, $N_{\text{t}} = 50$ and $N_{\text{u}} = 5$.} 
\label{fig:BER}       
\vspace{-1em}
\end{center}
\end{figure}
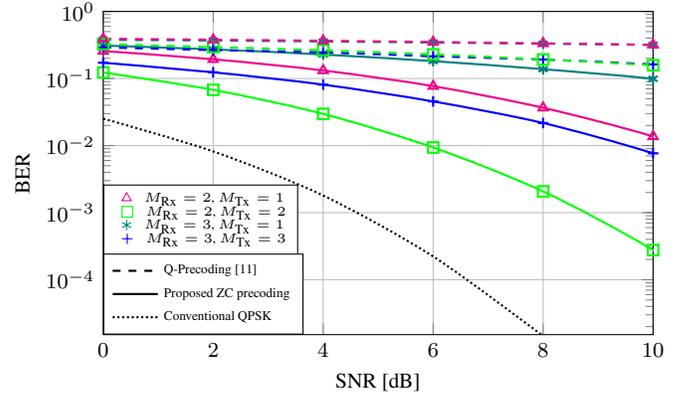
\begin{figure}[t]
\begin{center}
\input{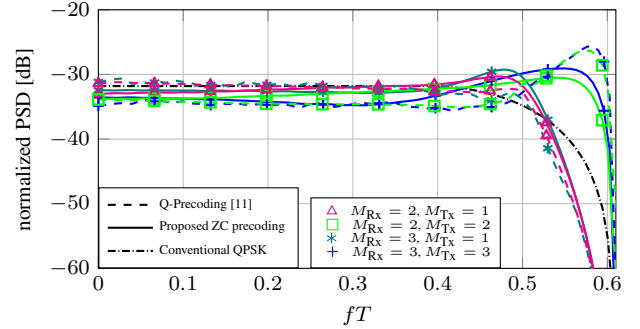}
\caption{Power spectral density for the experiment in Fig.~\ref{fig:BER}.} 
\label{fig:PSD}       
\vspace{-1em}
\end{center}
\end{figure}

\section{Conclusions}
\label{sec:conclusiones}

In this study we propose a precoding method for channels with 1-bit quantization and oversampling
at the receiver.
The novel optimization based precoding technique exploits the oversampling property in terms of conveying
 the information in the time instances of the zero-crossings of the received signal. 
Numerical results confirm a significantly larger distance to the decision threshold and superior BER performance in comparison with the state-of-the-art method.
At the same time the proposed approach obviates the need for dynamic codebook optimization, which reduces computational cost and saves band resources of the codebook transmission accordingly.

\newpage

\bibliographystyle{IEEEbib}
\bibliography{ref}

\end{document}